\documentclass[twoside,twocolumn,english]{revtex4-1}
\usepackage[T1]{fontenc}
\usepackage[utf8]{inputenc}
\usepackage[a4paper]{geometry}
\usepackage{amsmath}
\usepackage{graphicx}
\usepackage[charter]{mathdesign}
\makeatletter
\@ifundefined{date}{}{\date{}}
\makeatother

\usepackage{babel}
\begin{document}
\title{Spin density wave order in interacting type-I and type-II Weyl semimetals}
\author{Sarbajaya Kundu and David S\'en\'echal}
\affiliation{D\'epartement de physique and Institut quantique, Universit\'e de Sherbrooke,
Sherbrooke, Qu\'ebec, Canada J1K 2R1}
\date{\today}
\begin{abstract}
Weyl semimetals, featuring massless linearly dispersing chiral
fermions in three dimensions, provide an excellent platform for studying
the interplay of electronic interactions and topology, and exploring
new correlated states of matter. Here, we examine the effect of a
local repulsive interaction on an inversion-symmetry breaking Weyl
semimetal model, using cluster dynamical mean field theory
and variational cluster approximation methods. Our analysis
reveals a continuous transition from the gapless Weyl semimetal phase
to a gapped spin density wave ordered phase at a critical value
of the interaction, which is determined by the band structure parameters.
Further, we introduce a finite tilt in the linear dispersion and examine
the corresponding behavior for a type-II Weyl semimetal model,
where the critical interaction strength is found to be significantly
diminished, indicating a greater susceptibility towards interactions.
The behavior of different physical quantities, such as the double
occupancy, the spectral function and the Berry curvature associated
with the Weyl nodes are obtained in both the semimetallic and the
magnetically ordered states. Finally, we provide an interaction-induced
phase diagram for the Weyl semimetal model, as a function of the tilt
parameter. 
\end{abstract}
\maketitle

\section{\label{sec:introduction}introduction}

The interplay of electronic correlations and band topology
in strongly spin-orbit coupled systems \citep{doi:10.1146/annurev-conmatphys-020911-125138,doi:10.1146/annurev-conmatphys-031115-011319,Schaffer_2016}
has generated considerable interest in contemporary Condensed Matter Physics, with the potential to uncover new and exotic phases of matter.
In this context, topological insulators \citep{doi:10.1146/annurev-conmatphys-031214-014501,doi:10.1146/annurev-conmatphys-062910-140432,doi:10.7566/JPSJ.82.102001,Hasan_2015,RevModPhys.82.3045,RevModPhys.83.1057,Yan_2012}
and Weyl semimetals \citep{doi:10.1146/annurev-conmatphys-031016-025458,doi:10.1146/annurev-conmatphys-031113-133841,Jia2016,RevModPhys.90.015001}
are examples of strongly spin-orbit coupled systems with low-energy
degrees of freedom that are described by massless linearly dispersing
electrons, and are suitable for studying the combination of many-body
and band structure effects in topological materials. In particular,
Weyl semimetals (WSMs) feature pairs of nondegenerate bands touching
each other at isolated points in the band structure, with Weyl fermions
as low-energy quasiparticles. Theoretically, this can be realized by
breaking either inversion or time-reversal symmetry (TRS) or
both.
WSMs are characterized by open Fermi arcs~\citep{Deng2016,Huang2015,Jia2016,PhysRevLett.116.066802,Potter2014,Xu613,Xu_2015}
on their surfaces and a novel response to applied electric and magnetic
fields~\citep{Jia2016,PhysRevB.86.115133,PhysRevB.87.161107,PhysRevB.96.045112,PhysRevX.4.031035,PhysRevX.5.031023,Zhang2016}.
A conventional or type-I Weyl semimetal has a conical spectrum and a point-like
Fermi surface, but the energy dispersion at the node
could also be tilted along a given direction. When the tilt exceeds a certain critical value,
the Weyl node appears at the intersection of an electron and a
hole pocket, giving rise to a type-II WSM~\citep{Soluyanov2015,Deng2016,Jiang2017,PhysRevB.95.155124}.
The latter class of models are known to have properties qualitatively
different from those of the former, some of which include a field-selective
anomaly in magnetotransport~\citep{PhysRevLett.117.086401} and an intrinsic
anomalous Hall effect \citep{Zyuzin2016}. On the experimental side,
a number of material candidates for both type-I and type-II
Weyl semimetals have been proposed, and confirmed in recent times~
\citep{Deng2016,Jiang2017,PhysRevB.94.121113,Qi2016,Xu2015,Xu2016,Xu_2015,Xue1501092,Yang2015}.

Correlation effects are expected to be important for Weyl semimetal
candidates, which often involve heavier elements with a strong spin-orbit
interaction. Moreover, there have been experimental reports of collective
many-body effects, such as superconductivity \citep{Qi2016,LI2017}
or magnetism~\citep{Wang2018,PhysRevLett.124.017202}, in Weyl semimetals, necessitating
the theoretical treatment of electronic instabilities. Other important
questions include the robustness of the topological properties of
a Weyl semimetal in the presence of interactions~\citep{PhysRevB.100.245137,PhysRevB.94.125135,PhysRevB.94.241102,PhysRevB.96.195160,PhysRevB.98.241102,PhysRevLett.113.136402},
and the possibility of realizing interaction-induced topologically
nontrivial phases~\citep{PhysRevB.94.125135,PhysRevB.94.241102}. The effects of electronic interactions in WSMs have been explored
using various approaches \citep{Li2019exact,PhysRevB.100.245137,PhysRevB.89.014506,PhysRevB.89.235109,PhysRevB.90.035126,PhysRevB.94.125135,PhysRevB.94.241102,PhysRevB.95.201102,PhysRevB.96.045115,PhysRevB.96.165203,PhysRevB.96.195160,PhysRevB.98.241102,PhysRevLett.109.196403,PhysRevLett.113.136402,PhysRevResearch.2.012023,Yi2017possible,Zhang2018quantum,PhysRevB.97.125113,doi:10.7566/JPSJ.83.094710,PhysRevLett.122.046402}
such as perturbative renormalization group (RG)~\citep{PhysRevB.90.035126,PhysRevB.95.201102,PhysRevB.96.045115,Yi2017possible,Zhang2018quantum,PhysRevLett.124.127602}, mean-field analyses~\citep{PhysRevB.89.014506,PhysRevB.89.235109,PhysRevB.95.155124,PhysRevB.96.165203,PhysRevB.96.195160,PhysRevLett.109.196403,Berke_2018}, 
strong-coupling expansion methods~\citep{PhysRevB.94.125135} and, 
occasionally, numerical techniques~\citep{PhysRevB.94.241102,PhysRevLett.113.136402,PhysRevResearch.2.012023,PhysRevB.98.241102}.
Specific examples of possible broken-symmetry states have been considered,
such as excitonic and
charge-density wave (CDW) instabilities~\citep{PhysRevB.89.235109,PhysRevB.90.035126,PhysRevB.94.125135,PhysRevB.94.241102,PhysRevB.96.195160,PhysRevLett.109.196403,Yi2017possible}, 
as well as superconducting ground states~\citep{PhysRevB.89.014506,PhysRevB.95.155124,PhysRevB.93.094517,PhysRevB.92.035153}. 
There have been comparatively fewer studies of interaction
effects in type-II WSMs, generally using similar approaches~\citep{Yi2017possible,PhysRevB.96.165203,PhysRevB.97.161102}.
In order to complement the existing results, it is useful to employ
a nonperturbative approach, that can describe the
physical properties of the model by continuously varying the interaction
strength, producing results that are highly illustratory and comprehensive in nature. Besides, having a common framework to detect possible broken-symmetry
phases and examine the changes in the topological properties of the system, makes it easier
to characterize new and exotic types of order. Finally, one needs
a simple way of introducing a linear tilt in the dispersion, and examining
its effect on the properties of the interacting WSM model.

In this paper, we attempt to address these concerns,
by studying the effect of a local, repulsive interaction on a simple inversion-symmetry
breaking Weyl semimetal model, using two complementary methods, Cluster
Dynamical Mean Field Theory (CDMFT)~\citep{PhysRevLett.92.226402,PhysRevLett.101.186403,PhysRevB.91.235107,PhysRevB.62.R9283}
and the Variational Cluster Approximation (VCA)~\citep{PhysRevB.70.245110,PhysRevB.83.033104,PhysRevB.94.241102,PhysRevLett.91.206402,Shirakawa_2011}.
Both of these belong to a set of closely related approaches known
as Quantum Cluster Methods~\citep{Senechal2008qcm,PhysRevB.65.155112},
which consider a finite cluster of sites embedded in an infinite lattice,
and consider additional fields or ``bath'' degrees of freedom,
so as to best represent the effect of the surrounding infinite lattice.
The values of these additional parameters are decided using variational or self-consistency principles. In these approaches, broken symmetry
states can appear even for the smallest clusters used, and unlike
ordinary mean field theory, these are dynamical in nature, and retain
the full effect of strong correlations. These methods allow us to
obtain the full interacting Green's function, spectral functions and
topological properties as a function of the interaction strength,
and include additional tuning parameters, such as a finite tilt in
the dispersion, with relative ease. 

\begin{figure}
\begin{centering}
\includegraphics[width=1\columnwidth]{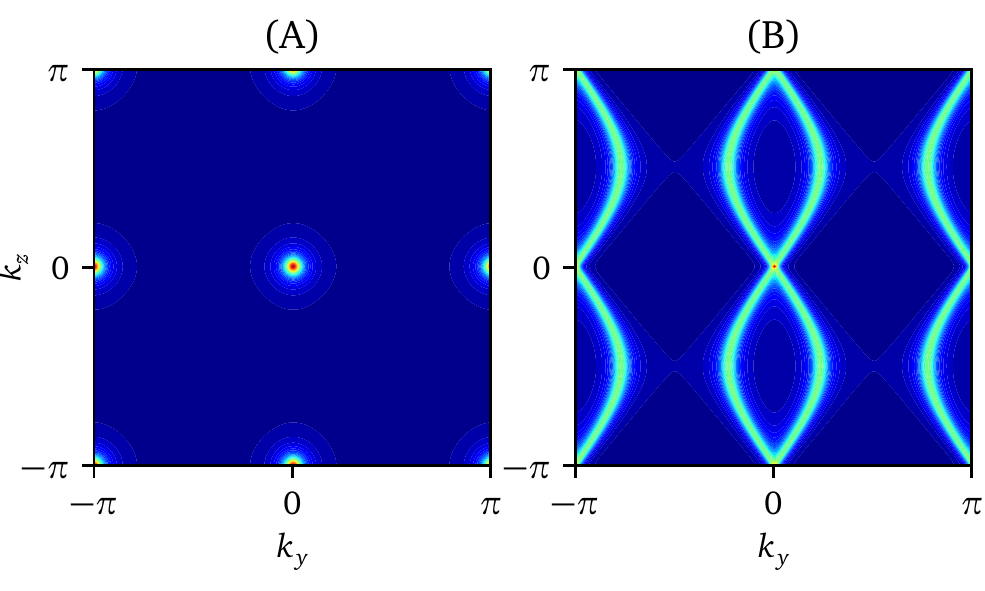}
\par\end{centering}
\caption{\label{fig:mdc}Projection of the Fermi surface on the $k_{x}=0$
plane, for the WSM model defined in Eq.~\eqref{eq:1-1}, with band structure
parameters $v_{x}=v_{y}=v_{z}=1$, and (A) $w_{z}=0$ (type-I) for $U=1$,
and (B) $w_{z}=1.2$ (type-II) for $U=0$. In the former
case, we observe Fermi points at the positions of the Weyl
nodes in the plane, i.e. $(0,0,0)$, $(0,0,\pm\pi)$, $(0,\pm\pi,0)$
and $(0,\pm\pi,\pm\pi)$, whereas in the latter case, each Fermi point
is replaced by an electron and a hole pocket, connected by the corresponding
Weyl node, the boundaries of which can be clearly seen along the $k_{z}$-direction.
Note that we choose an interaction strength $U<U_{c}$ for the above
illustration, where $U_{c}$ is a critical value of the interaction
amplitude, at which a gap opens in the spectrum and the system becomes
magnetically ordered. }
\end{figure}

Our main findings are as follows. At a critical value of the interaction
strength, $U=U_{c}$, the Weyl semimetal undergoes a continuous transition
to a topologically trivial spin density wave (SDW) ordered state.
We find an ordering wavevector $Q=(0,0,\pi)$, which connects Weyl
nodes of opposite chiralities, with the magnetization pointing in
the $z$ direction. In the rest of the paper, we denote this particular
SDW order as $M_{z}^{(0,0,\pi)}$. For the untilted Weyl dispersion,
it is equivalent to a state with an ordering wavevector $Q=(\pi,0,0)$
(see Appendix A), with the magnetization pointing in the $x$ direction,
which we henceforth denote as $M_{x}^{(\pi,0,0)}$. 
These particular orders are presumably
favored by the presence of nesting in the band structure, between Weyl
nodes of opposite chiralities. Since our analysis is limited to the
effect of repulsive local interactions at half-filling, we constrain
our attention to various spin density wave instabilities. As expected,
the magnetic order is accompanied by the gradual appearance of a spectral
gap, and the destruction of the Berry curvature associated with the
Weyl nodes, which we compute for the folded band structure, in the ordered state. As the
interaction amplitude increases, the SDW order is found to become
more robust, as indicated by a gradual increase in the magnitude of
the order parameter. The critical interaction $U_{c}$, at which it
appears, also depends on the details of the band structure and the
model parameters considered. We introduce a finite tilt parameter
$w_{z}$ (assumed to be along the $k_{z}$ direction for simplicity)
and find that in the over-tilted type-II regime, the transition
occurs at a significantly diminished value of the interaction, $U_{c}$.
This is consistent with the expectation of an increased sensitivity
towards interactions, when each Fermi point is replaced by an electron
and a hole pocket (see Fig.\ref{fig:mdc}(B)). Once again, the SDW
order $M_{z}^{(0,0,\pi)}$ (which is no longer equivalent to $M_{x}^{(\pi,0,0)}$
for $w_{z}\neq0$) is found to be favored for the type-II WSM
model. In general, the magnitude of the tilt parameter strongly affects the critical interaction amplitude for the transition, and its orientation may decide the specific nature of the SDW order. In particular, a different
spin density wave order denoted as $M_{z}^{(\pi,\pi,\pi)}$, with
an ordering wavevector $Q=(\pi,\pi,\pi)$, is also found to compete with
 $M_{z}^{(0,0,\pi)}$, for a more general direction of the tilt (see Appendix A
for a pictorial depiction of the different types of magnetic order
that can be realized in this system). We have
independently verified our results for the order of the phase transition
and the competing density-wave instabilities, using both the CDMFT
and the VCA approaches. 

The paper is organized as follows. In Sec.~\ref{sec:model} we introduce
the model Hamiltonian, and provide a brief overview of the CDMFT and
VCA methods that are used in our analysis. In Sec.~\ref{sec:results}
we present the results of our CDMFT computations for the double occupancy
and the dominant SDW order parameter, as a function of the interaction.
We also pictorially illustrate the behavior of the spectral function
and the Berry curvature associated with the Weyl nodes. We then present
the results of our VCA calculations, which confirm the nature of the
spin density wave instability occuring in this system, as well as the
order of the transition. We also present the corresponding results
as a function of the increasing tilt parameter $w_{z}$, which pushes
the critical interaction strength $U_{c}$ to smaller values. Based
on these results, we present the interaction-induced phase diagram
of the WSM model considered by us, as a function of $w_{z}$. Finally,
in Sec.~\ref{sec:conclusions}, we summarize our results, discuss
some relevant observations, and present the conclusions of our study.

\section{\label{sec:model}model and methods}

\subsection{Model Hamiltonian}

\begin{figure}
\begin{centering}
\includegraphics[width=0.8\columnwidth]{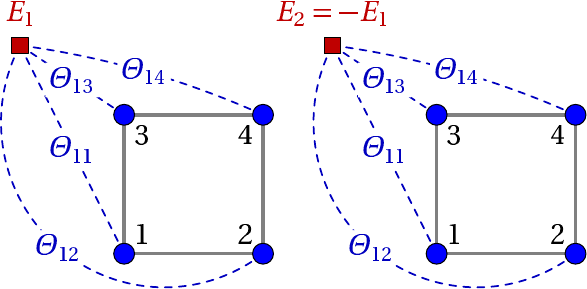}
\par\end{centering}
\caption{\label{fig:bath}A part of the 12-site $(4\times8)$ cluster-bath
system used in our CDMFT computations, with 4 cluster sites and 8
bath sites in total, depicting one of the bath sites. For model parameters
in the type-I regime, with $|w_{z}|<1$, we use a particle-hole
symmetric ansatz for the bath parameters, and each cluster site is
hybridized with every bath site in the system. This situation is pictorially
depicted here. The bath energies are denoted by $E_{i}(i=1-8)$, and
these are hybridized with the cluster sites with hopping amplitudes
$\Theta_{ir}(i=1-8,r=1-4)$. We consider $E_{2}=-E_{1}$
in this case, as shown above.}
\end{figure}

Consider the general form of the Hamiltonian in the vicinity of a
Weyl point, along with a possible tilt in the dispersion, given by
\begin{equation}
H_{0}(k)=w_{z}k_{z}+v_{x}k_{x}\sigma_{x}+v_{y}k_{y}\sigma_{y}+v_{z}k_{z}\sigma_{z},\label{eq:1}
\end{equation}
where $\sigma_{x,y,z}$
are the Pauli matrices. The corresponding eigenvalues are 
\[
E_{\pm}(k)=w_{z}k_{z}\pm\sqrt{v_{x}^{2}k_{x}^{2}+v_{y}^{2}k_{y}^{2}+v_{z}^{2}k_{z}^{2}}.
\] 
The first term is a linear tilting term, which is crucial for the
emergence of type-II Weyl fermions. Once $|w_{z}|>|v_{z}|$, the Weyl cone is over-tilted and
the Fermi surface changes from a point to an electron and a hole pocket,
touching each other at the type-II Weyl node. Throughout our
analysis, we consider $v_{x}=v_{y}=v_{z}=1$ and tune the tilt parameter $w_{z}$,
to explore different regimes. When $|w_{z}|<|v_{z}|=1$, we have a
conventional type-I Weyl semimetal, while $|w_{z}|>1$ corresponds
to the over-tilted type-II regime. In principle, one could also
have a non-linear dispersion in one or more directions in Eq.~\eqref{eq:1}
above, and include a quadratic tilt term ( referred to as a Type-III WSM \cite{Li2019Type} ). This model has its own
peculiar properties and will be studied in a future work. 

For the purpose of our analysis, we study models defined on the cubic
lattice. Using Eq.~\eqref{eq:1} above, we consider the following noninteracting
lattice model for a Weyl semimetal with a linear dispersion, 
\begin{multline}
H_{0} =\sum_{k}c_{k}^{\dagger}\Big(w_{z}\sin k_{z}+v_{z}\sigma_{z}\sin k_{z}\\
+v_{x}\sigma_{x}\sin k_{x}+v_{y}\sigma_{y}\sin k_{y}\Big)c_{k},\label{eq:1-1}
\end{multline}
where the lattice spacing has been set to unity, and the fermion operators
are spin doublets, i.e. $c_{k}\equiv(c_{k\uparrow},c_{k\downarrow})$.
The Weyl nodes for the above model occur at the points $(0,0,0)$, $(\pi,0,0)$, $(\pi,\pi,0)$, $(\pi,\pi,\pi)$ and permutations thereof. 
One of the consequences of the presence of Weyl nodes at these high-symmetry points is the absence of Fermi arcs in the usual orthogonal $x-$, $y-$ or $z-$directions. This may be due to the fact that the projected Weyl points on the surfaces in all of these directions are formed by Weyl points of opposite chiralities (see Fig. 5). However, we do see Fermi arcs in the (110) direction for this model. 
When $w_{z}=0$, the above model preserves particle-hole and time-reversal
symmetry, as well as $C_{4}$ rotational symmetry about the $x-$, $y-$ and $z-$axes, but breaks inversion symmetry. In the presence of a tilt along the $z-$direction, the model continues to preserve particle-hole symmetry, along with $C_{4}$ rotational symmetry with respect to the $z-$axis. It should be noted that in the presence of the tilt $w_{z}$, we use the particle-hole symmetry operation $c_{k\sigma}\rightarrow U_{\sigma\sigma^{\prime}} c^{\dagger}_{k \sigma^{\prime}}$ with $U=i \sigma_{z}$, leading to $H_{0}(k)\rightarrow-(U^{\dagger}H_{0}(-k)U)^{*}$, augmented by a reflection $k_{x} \rightarrow -k_{x}$. A similar transformation is also required for the untilted WSM model.

In the real-space representation, the noninteracting Hamiltonian takes
the following form
\begin{multline}
H_{0} = \frac{-i}{2}\sum_{r,\alpha\beta}\Big(w_{z}c_{r\alpha}^{\dagger}c_{r+\hat{z}\beta}+v_{z}c_{r\alpha}^{\dagger}\sigma^{z}_{\alpha\beta}c_{r+\hat{z}\beta} \\
+v_{x}c_{r\alpha}^{\dagger}\sigma^{x}_{\alpha\beta}c_{r+\hat{x}\beta}+v_{y}c_{r\alpha}^{\dagger}\sigma^{y}_{\alpha\beta}c_{r+\hat{y}\beta}\Big)-\mu\sum_{r,\sigma}n_{r,\sigma},\label{eq:2}
\end{multline}
Here, $\hat{x}$, $\hat{y}$ and $\hat{z}$ are the lattice unit vectors
along the $x$, $y$ and $z$ directions, the operator $c_{r\alpha}$
annihilates a particle with spin $\alpha$ on site $r$, while $\sigma^{\mu}$
($\mu=x,y,z$) denotes the three Pauli matrices corresponding to the
spin degree of freedom, and the number
density per spin projection of the spin-1/2 electrons is $n_{r,\sigma}=c_{r,\sigma}^{\dagger}c_{r,\sigma}$. In the following analysis, we investigate
the effect of local Hubbard interactions on the model defined above.
The resulting Hamiltonian is as follows,
\begin{equation}
H=H_{0}+U\sum_{r}n_{r,\uparrow}n_{r,\downarrow},\label{eq:4}
\end{equation}
where $H_{0}$ is defined in Eq.~\eqref{eq:2} above.
$U$ is the Hubbard interaction parameter which is taken to be positive,
or repulsive. For the purpose of our analysis, the chemical potential
is fixed at $\mu=U/2$ throughout, which corresponds to half-filling.

As mentioned earlier, we examine the possibility of spin density wave
(SDW) orders as prototypical many-body instabilities of the WSM model
at half-filling with repulsive interactions\footnote{While superconductivity and charge-density wave order are ruled out due to these conditions, we have also examined the possibility of on-site and bond ferromagnetism, which are not found to occur in this model.}. A general SDW operator
with wavevector $Q$ is defined as follows-
\begin{equation}
\Psi_{SDW}=\lambda\sum_{r}A_{r}\cos[Q.r+\phi],\label{eq:5}
\end{equation}
where $A_{r}=S_{r}^{z},S_{r}^{x}$, and $\lambda$ is a variational
parameter. In the following analysis, we probe the presence of SDW
orders with wavevectors $Q=(\pi,0,0)$, $(0,0,\pi)$, $(\pi,\pi,0)$
and $(\pi,\pi,\pi)$, and
observe stable solutions for different values of $Q$ in the strongly
interacting regime, depending on the parameters chosen.

\subsection{Methods: CDMFT and VCA}

Here, we provide a brief overview of the Quantum Cluster Methods used
in our analysis. For a more detailed discussion on the principles
and the mathematical background of these methods, please see
Ref. \onlinecite{Senechal2008qcm}. 

Cluster Dynamical Mean Field Theory (CDMFT) is an extension of the
Dynamical Mean Field Theory (DMFT) method, where instead of a single-site impurity,
we consider a cluster of sites with open boundary conditions,
taking into account short-range spatial correlations exactly.
In this approach,  a set of uncorrelated,
additional ``bath'' orbitals hybridized with the cluster are used to account for the effect of the cluster's environment.
Thus, the infinite lattice is tiled into small clusters, and each of these
is coupled to a bath of uncorrelated, auxiliary orbitals. These
bath orbitals have their own energy levels
$E_{i\sigma}$, which may be spin-dependent, and are hybridized with the cluster sites
$r$ with amplitudes $\Theta_{ir\sigma}$. The bath parameters $(E_{i\sigma},\Theta_{ir\sigma})$
are determined by a self-consistency condition (see Ref. \onlinecite{Senechal2008qcm} for
details). 

In our analysis, we use a 12-site ($4\times8$) cluster-bath system,
a part of which is illustrated in Fig. \ref{fig:bath}. An effective model is
solved on the cluster, and the self-energy associated with that cluster
is then applied to the whole lattice. The lattice Green's function is
computed from the cluster's self-energy $\Sigma(\omega)$ as 
\[
G^{-1}(\tilde k,\omega)=G_{0}^{-1}(\tilde k,\omega)-\Sigma(\omega)
\]
where $\tilde k$ denotes a reduced wavevector (defined in the
Brillouin zone of the super-lattice), and $G_{0}$ is the noninteracting
Green's function. Once a solution is found for a given set of model
parameters, the above Green's function $G$ can be used to obtain the average values of one-body operators defined on the lattice. An
exact diagonalization solver is used (at zero temperature), and the computational size of the problem is determined by the total number of cluster and bath orbitals. For the type-I
WSM model with $|w_{z}|<1$, we consider a particle-hole symmetric
ansatz for the bath parameters, and every cluster site is hybridized
with every bath site in the system by a hopping parameter. On the other hand, for the
type-II WSM model with $|w_{z}|>1$, each cluster site is only hybridized with the
two bath sites that are adjacent to it, for simplicity. We do not assume other symmetries of the model, such as the $C_{4}$ rotational symmetry, during our analysis. 

The Variational Cluster Approximation (VCA) method involves solving
a model exactly on a small cluster of lattice sites, after inserting fields
that represent the effect of the cluster's environment. The essence
of this method lies in Potthoff's self-energy functional approach
(SFA) \cite{Potthoff2011}, involving a functional $\Omega[\Sigma]$ of the self-energy
$\Sigma$, that is parametrized by the one-body terms collectively labeled
by $h$. The original Hamiltonian $H$, defined on the infinite lattice,
is considered along with a reference Hamiltonian $H^{\prime}$, which
is often a restriction of $H$ to the cluster. In order to probe different
broken symmetries, a finite number of
Weiss fields may be added to the latter. Any one-body term can be added to $H^{\prime}$,
since the basic requirement is that $H^{\prime}$ and $H$ must share
the same interaction term. The electron self-energy $\Sigma(\omega)$
associated with $H^{\prime}$ is then used as a variational self-energy to construct the Potthoff self-energy functional (see Ref.
 \onlinecite{Senechal2008qcm}
for a detailed mathematical explanation of the VCA approach) :
\begin{align}
\Omega[\Sigma(h)] & =\Omega^{\prime}[\Sigma(h)]+\mathrm{Tr}\ln[-(G_{0}^{-1}-\Sigma(h))^{-1}]\nonumber \\
 & -\mathrm{Tr}\ln(-G^{\prime}(h)),\label{eq:11}
\end{align}
where $G^{\prime}$ is the physical Green's function of the cluster,
$G_{0}$ is the noninteracting Green's function of the original lattice
model, and $h$ jointly denotes the coefficients of the adjustable
one-body terms added to $H^{\prime}$ which act as variational parameters.
The symbol $\mathrm{Tr}$ stands for a sum
over all degrees of freedom and frequencies. $\Omega^{\prime}$ refers to
the ground-state energy of the cluster which, along with the associated
Green's function $G^{\prime}$, is computed
via the exact diagonalization method (at zero temperature) in our case. The stationary
point of the functional gives the best
possible self-energy $\Sigma(\omega)$. This is combined with $G_{0}$
to form an approximate Green's function $G$ for the original Hamiltonian
$H$, from which any one-body term, such as the order parameters associated
with various SDW orders, can be computed. We define our reference
Hamiltonian on an eight-site cubic cluster. 

\section{\label{sec:results}results and discussion}

In this section, we discuss the behavior of different physical quantities
associated with the WSM model defined in Eq.~\eqref{eq:1-1}, as a function
of the interaction strength $U$, obtained from our analysis. Using
CDMFT, we examine the double occupancy $d= \langle n_{\uparrow} n_{\downarrow} \rangle$, the relevant SDW order parameters,
the spectral function and the Berry curvature associated with the
Weyl nodes, for both the type-I and type-II WSM models.
In the VCA approach, we first obtain the behavior of the Potthoff
functional $\Omega$ as a function of the relevant Weiss fields, corresponding
to different types of order. The stationary point of the functional
is then used to approximately evaluate the Green's function for the
lattice model, and calculate different physical properties as a function
of the interaction amplitude. We also compare the critical interaction
strength for different values
of the tilt parameter $w_{z}$. Finally, we obtain the interaction-induced
phase diagram of the system, as a function of $w_{z}$. 

\subsection{CDMFT: Numerical results}

\begin{figure}
\begin{centering}
\includegraphics[width=0.9\columnwidth]{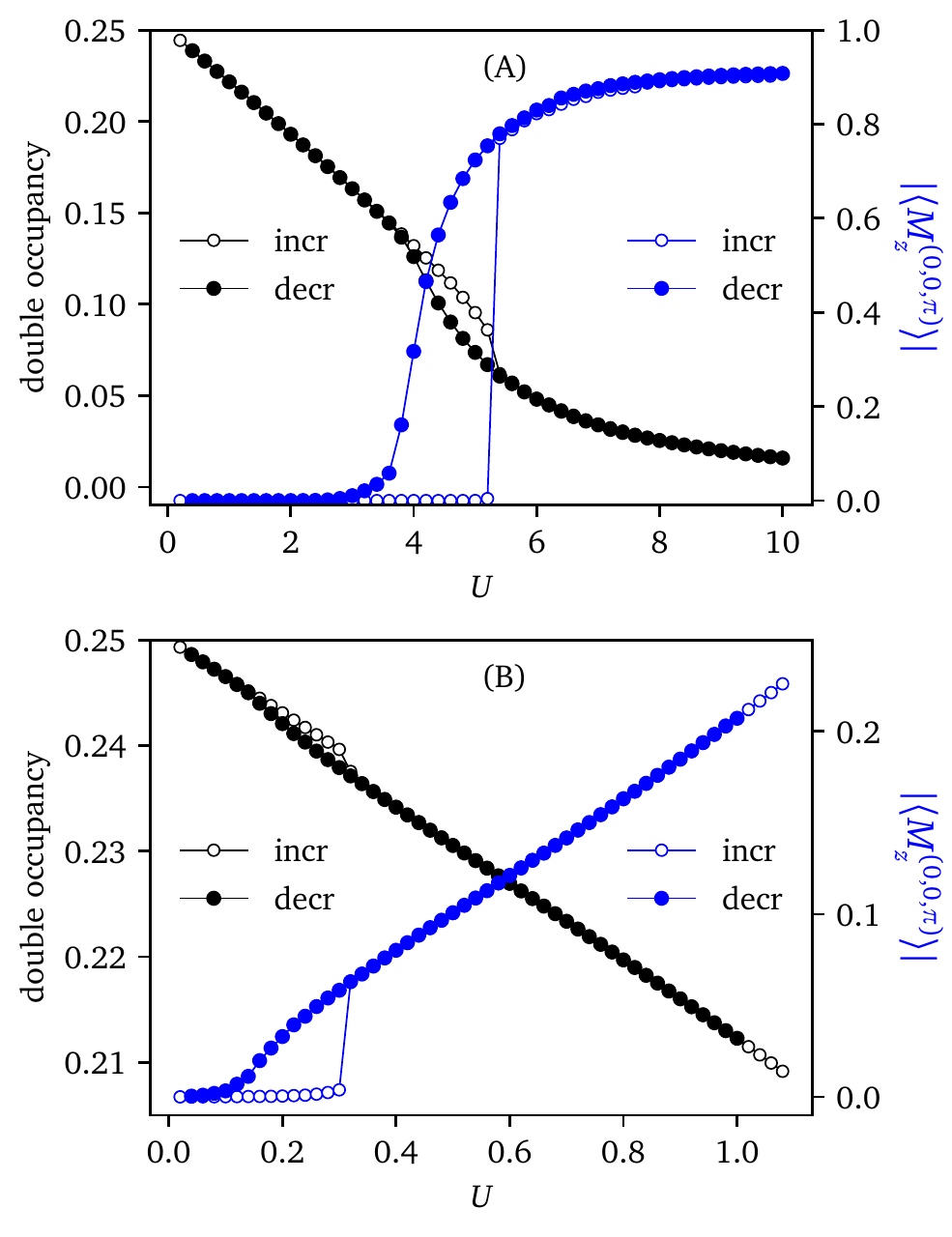}
\par\end{centering}
\caption{\label{fig:doubleocc_order}Evolution of the double occupancy and
the spin density wave (SDW) order parameter $\langle M{}_{z}^{(0,0,\pi)}\rangle$
with an ordering wavevector $Q=(0,0,\pi)$, as a function of the interaction
parameter $U$ for two sets of model parameters (A) $v_{x}=v_{y}=v_{z}=1$, $w_{z}=0$
(type-I, above) and (B) $v_{x}=v_{y}=v_{z}=1$, $w_{z}=1.5$ (type-II,
below). For the untilted dispersion, this is equivalent to the order
$M_{x}^{(\pi,0,0)}$, with $Q=(\pi,0,0)$. In both the cases, we observe
a continuous transition to an SDW ordered state, for a critical value
of the interaction strength, $U_{c}$. For $w_{z}=0$, we find $U_{c}\approx3$,
whereas for $w_{z}=1.5>v_{z}$, the system reacts much more strongly
to the presence of interactions, with $U_{c}\sim0.1$. This is expected
due to the appearance of a finite Fermi surface for an over-tilted
dispersion. As explained in the text, we find a jump in the order
parameter $\langle M{}_{z}^{(0,0,\pi)}\rangle$ as a function of increasing
$U$ in our CDMFT calculations for a value of $U>U_{c}$, which leads
to an apparent hysteresis behavior. However, as we decrease $U$,
the system seems to undergo a continuous transition at $U=U_{c}$.}
\end{figure}

\begin{figure}
\begin{centering}
\includegraphics[width=1\columnwidth]{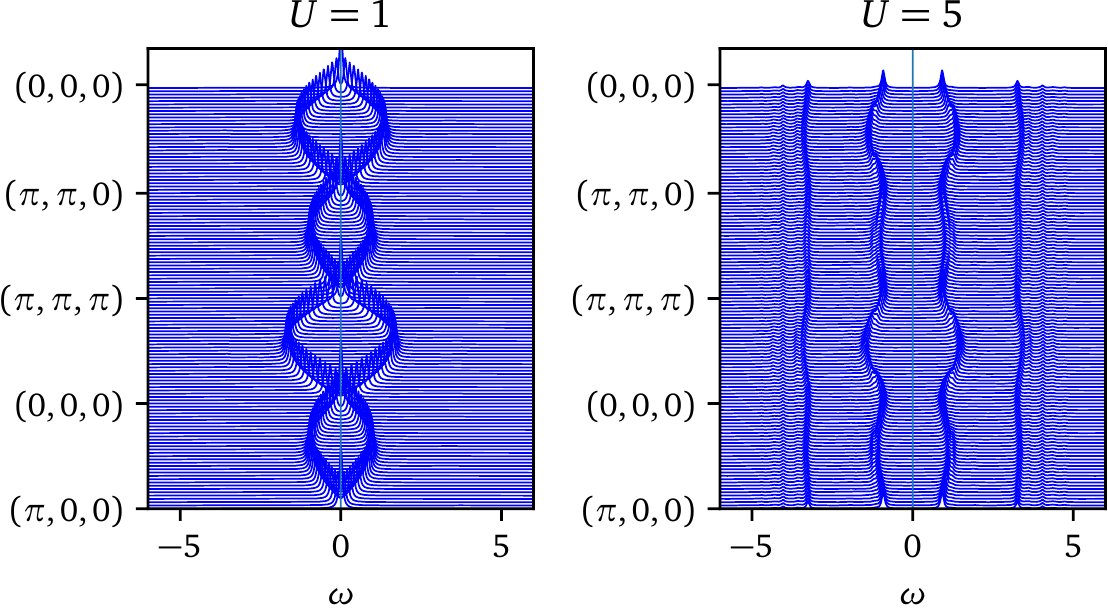}
\par\end{centering}
\begin{centering}
\includegraphics[width=1\columnwidth]{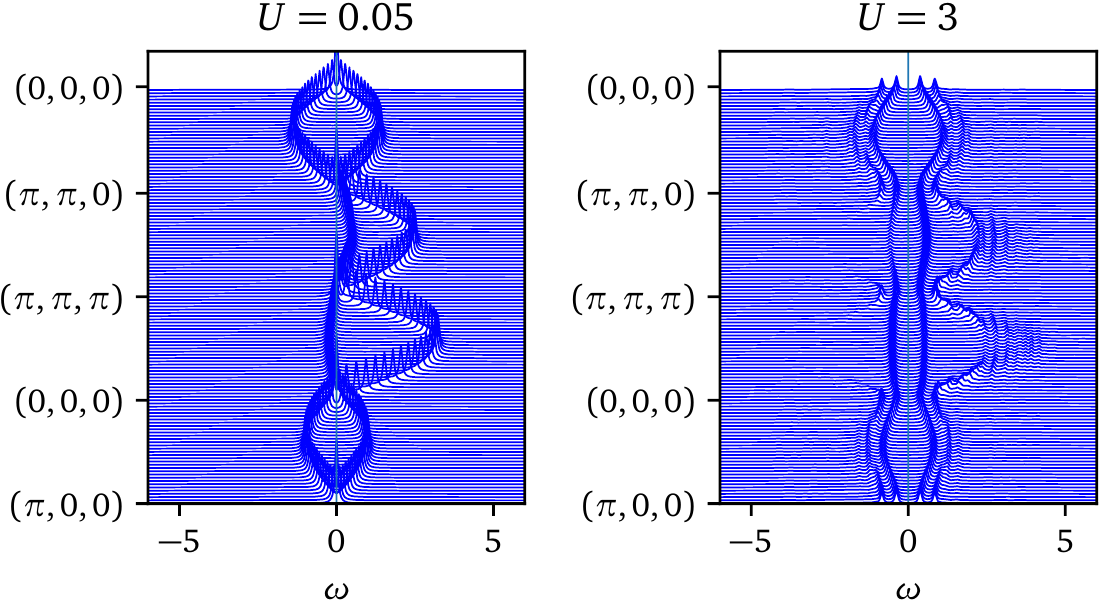}
\par\end{centering}
\caption{\label{fig:spectral1}The spectral function, obtained from our CDMFT
calculations, for the gapless Weyl semimetal phase, as well as the
magnetically ordered phase. The panels above and below correspond
to the type-I and type-II regime of parameters, respectively.
In the upper panel, we present the spectral functions for the band structure
parameters $v_{x}=v_{y}=v_{z}=1$ and $w_{z}=0$, considering $U=1<U_{c}$,
where the system is still gapless, and $U=5>U_{c}$, where a finite
gap has appeared in the spectrum. The lower panel shows the corresponding
results for $w_{z}=1.5$, with the interaction amplitudes $U=0.05<U_{c}$
and $U=3>U_{c}$. From Fig.~\ref{fig:doubleocc_order}(B), we observe
that the order parameter $\langle M{}_{z}^{(0,0,\pi)}\rangle$ varies
very slowly as a function of $U$ in the over-tilted regime, which
may lead to a more gradual appearance of the spectral gap.}
\end{figure}

\begin{figure}
\begin{centering}
\includegraphics[scale=0.9]{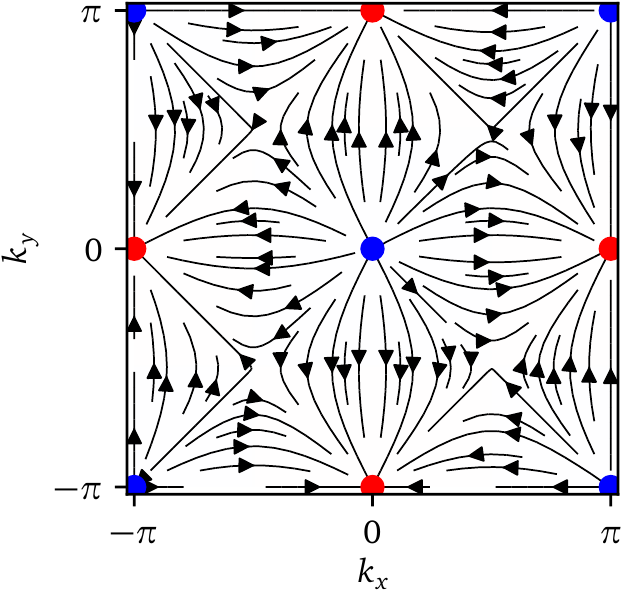}
\par\end{centering}
\caption{\label{fig:berry}The field lines of the Berry curvature $\nabla\times A$
in the $k_{z}=0$ plane, for the type-I WSM model with parameters
$v_{x}=v_{y}=v_{z}=1$ and $w_{z}=0$, considering $U=2<U_{c}$, where the system
is still in the semimetallic state. As expected, we obtain the same
result in the presence of a finite $w_{z}$. In
this plane, the Weyl nodes are present at the points $(0,0,0)$, $(\pm\pi,0,0)$,
$(0,\pm\pi,0)$ and $(\pm\pi,\pm\pi,0)$. The directions of the arrows and the colors (red or blue)
indicate the chiralities associated with the different nodes. At a
critical interaction strength $U_{c}$, we observe a transition to
an SDW order $M_{z}^{(0,0,\pi)}$, with an ordering wavevector $Q=(0,0,\pi)$
(or, equivalently $M_{x}^{(\pi,0,0)}$ with $Q=(\pi,0,0)$ for $w_{z}=0$) which
connects Weyl nodes of opposite chiralities. In the ordered state,
backfolding due to the ordering wavevector $Q$ maps these Weyl nodes
to one another, leading to a gapped spectrum, and the Berry curvature
vanishes. }
\end{figure}

\subsubsection{Type-I:}

Here, we discuss our CDMFT results for a type-I Weyl semimetal
model, with $v_{x}=v_{y}=v_{z}=1$ and $w_{z}=0$. These results remain qualitatively
unchanged for all $|w_{z}|<|v_{z}|$. 

\paragraph{Double occupancy and order parameter:}

For the above values of the band structure parameters, our
CDMFT solutions indicate that the Weyl semimetal undergoes a continuous
transition to the spin-density wave (SDW) ordered state $M_{z}^{(0,0,\pi)} \propto \sum_{r}S_{r}^{z}\cos[Q.r+\phi]$ with ordering wavevector $Q=(0,0,\pi)$ and the magnetization pointing in the $z-$direction,
at a critical interaction amplitude $U=U_{c}\approx3$. We find clear
signatures of a phase transition in the double occupancy, as well
as the magnitude of the SDW order parameter $|\langle M{}_{z}^{(0,0,\pi)}\rangle|$
calculated as a function of $U$ (see Fig.~\ref{fig:doubleocc_order}).
However, there is an apparent discrepancy between the behavior of
these quantities for increasing and decreasing values of $U$. In
the former case, we observe a jump in the double occupancy as well
as the order parameter $|\langle M{}_{z}^{(0,0,\pi)}\rangle|$, for
$U\approx5>U_{c}$. This is evident from Fig.~\ref{fig:doubleocc_order}(A).
However, as we decrease the magnitude of $U$, no such jump is observed,
and the transition appears to be continuous. Therefore, an apparent
hysteresis behavior is observed in the transition region of $U_{c1}<U<U_{c2}$
where $U_{c1}\approx3$ and $U_{c2}\approx5$, which also shows up
in the double occupancy. The reason for observing a finite jump in
the order parameter is unclear, but it could potentially be due to
a false minimum generated in the self-consistency procedure employed
in CDMFT, that depends on the direction of the change in $U$. In the next section, we find that our VCA results clearly
point towards a second-order phase transition in this system. 

\paragraph{Spectral function:}

We also calculate the spectral function $A(\omega,k_{x},k_{y},k_{z})=-\frac{1}{\pi}ImG(\omega,k_{x},k_{y},k_{z})$
for the WSM model, and illustrate our results for two different values
of the interaction parameter $U$, representative of the semimetallic
and $M_{z}^{(0,0,\pi)}$ phases respectively (see the upper panel in Fig.~\ref{fig:spectral1}).
At $U=1<U_{c}$, the spectrum is found to be gapless, as expected,
and the dispersion resembles that of a noninteracting type-I
WSM. At $U=5>U_{c}$, a large spectral gap is observed, along with
a nontrivial value for the SDW order parameter $\langle M{}_{z}^{(0,0,\pi)}\rangle$.
This is consistent with the expectation of a phase transition at $U=U_{c}\approx3$. 

\paragraph{Topological properties:}

In order to verify the topological properties of the WSM model in
the presence of interactions, we calculate the Berry phases associated
with the Weyl nodes, using an approach introduced in Ref. \onlinecite{PhysRevLett.113.136402},
which we briefly describe below. 

In a noninteracting WSM, the Weyl points can be identified as
hedgehog singularities of the Berry curvature, $\nabla\times a(k)$, where $a$ is the Berry connection defined in terms of the occupied Bloch
states. In the presence of interactions, a many-body Berry
connection $A(k)$ and associated Berry curvature $\nabla\times A$ are analogously defined in Ref. \onlinecite{PhysRevLett.113.136402},
using the zero-frequency Green's function. A topological
Hamiltonian is defined as,
\[
H_{t}(k)=-G(0,k)^{-1}=H(k)+\Sigma(0,k)
\]
where $H$ is the Bloch Hamiltonian for the noninteracting case,
while $\Sigma(i\omega,k)$ is the self-energy matrix. $H_{t}$
plays the role of an effective Bloch Hamiltonian for the interacting system, and its eigenstates are used to define the many-body Berry connection, as $A(k)=-i\sum\langle nk|\nabla|nk\rangle$,
where $H_{t}(k)|nk\rangle=\widetilde{\epsilon_{n}}(k)|nk\rangle$, $\{\widetilde{\epsilon_{n}}(k)\}$
being the band structure of $H_{t}$. Here the sum is restricted to eigenstates with $\widetilde{\epsilon_{n}}(k)\le0$. It has been argued that the monopoles of $\nabla\times A$ correspond to the Weyl points of the interacting system. 

Using the above approach, we have calculated the Berry curvature for
different band structure parameters, and a range of values of $U$. Fig.~\ref{fig:berry} shows the field lines of $\nabla\times A$ in
the $k_{z}=0$ plane, for $U=2<U_{c}$, which illustrates the topological properties of the WSM model in the presence of interactions. 
In the ordered state, we consider a folded Brillouin zone for calculating
the Berry curvature, which is found to vanish. Backfolding due to
an ordering vector $Q$ maps Weyl nodes to regions of the Brillouin
Zone with nodes of opposite chirality, where they meet and gap out. We have also verified that the Fermi arcs in this system vanish in the gapped state.

\subsubsection{Type-II (over-tilted):}

Next, we consider the WSM model in the over-tilted type-II regime,
i.e. when the tilt parameter $|w_{z}|>|v_{z}|$. As an illustrative
example, we discuss our results for the band structure parameters $v_{x}=v_{y}=v_{z}=1$
and $w_{z}=1.5$. 

\paragraph{Double occupancy and order parameter:}

When the tilt $w_{z}$ exceeds a critical value, the Fermi points
are replaced by electron and hole pockets touching each other at the
type-II Weyl node, the outlines of which can be seen on the
$k_{x}=0$ plane, along the $k_{z}$ direction, in Fig.~\ref{fig:mdc}(B).
Due to the presence of a finite density of states at the Fermi level
in this case, the WSM phase is expected to be more susceptible to interaction
effects, which is confirmed in our analysis. Here, the system again
undergoes a second-order transition to the SDW ordered state $M_{z}^{(0,0,\pi)}$
(which is no longer equivalent to $M_{x}^{(\pi,0,0)}$ for $w_{z}\neq0$)
at a critical value of the interaction $U=U_{c}$, where $U_{c}\sim0.1$.
Note that this is significantly diminished in comparison to the previous
case, where $U_{c}\approx3$. We find that for increasing $U$, the
double occupancy as well as the order parameter $\langle M{}_{z}^{(0,0,\pi)}\rangle$
show a small jump at around $U\sim0.3$, and an apparent hysteresis
behavior is again observed in the region $U_{c1}=0.1<U<U_{c2}=0.3$.
However, it is found to be less prominent than in the case of a type-I
WSM. As we decrease the magnitude of $U$, no such jump is observed
and the transition is appears to be continuous (see Fig.~\ref{fig:doubleocc_order}(B)). 

\paragraph{Spectral function:}

We also show the spectral function obtained in this case for two representative
values of $U$, i.e. $U=0.05<U_{c}$ and $U=3>U_{c}$, in the lower panel in Fig.~\ref{fig:spectral1}.
In this case, we find a relatively gradual opening of the spectral
gap, as compared to the upper panel with $w_{z}=0$,
which may be due to a much slower variation in the magnitude of the
order parameter $\langle M{}_{z}^{(0,0,\pi)}\rangle$ in this case.

To conclude this part of our analysis, our CDMFT solutions for the
WSM model in Eq.~\eqref{eq:1-1}, with interactions, indicate a second-order
phase transition to a spin density wave ordered state at a critical
interaction strength $U=U_{c}$. This is accompanied by the appearance
of a gap in the spectral function and the vanishing of the Berry curvature
associated with the Weyl nodes. For both the type-I and type-II
WSM models considered above, we observe an apparent jump in the magnitude
of the order parameter for increasing values of $U$, which is absent
for decreasing $U$. This could be due to a false minimum in the CDMFT
procedure, which is prone to first-order transitions. The SDW order
is found to become more robust for larger values of $U$, as indicated
by a slow increase in the magnitude of the order parameter for $U>U_{c}$.
In the next section, we confirm the order of the transition using
the VCA approach. 

\subsection{VCA: Numerical results}

We have also used the Variational Cluster Approximation (VCA) method
to investigate the effect of local repulsive interactions on the WSM
model. As stated earlier, we limit our considerations to spin density
wave instabilities, and probe the relevant SDW orders by solving the
following Hamiltonian on an 8-site cubic cluster, 
\[
H^{\prime}=H_{0}^{\prime}+ h M_{z}^{(0,0,\pi)\prime}
\]
where $H_{0}^{\prime}$ and $M_{z}^{(0,0,\pi)\prime}$ are the
restriction to the cluster of the kinetic energy operator, and the
SDW operator $M_{z}^{(0,0,\pi)}$ defined in Eq.~\eqref{eq:5}, respectively.
The coefficient $h$ is the corresponding
Weiss field, which is the variational parameter used in optimizing
the Potthoff functional $\Omega$. We study the evolution of the Potthoff
functional $\Omega$, as a function of $h$,
for different values of $U$. Initially, for a weakly interacting
system, $\Omega$ has a single minimum at $h=0$ ,
indicating the absence of the corresponding SDW order. At a critical
value $U=U_{c}$, it develops a new minimum at a finite value of the
Weiss field $h$ , along with a maximum
at $h=0$  and this behavior of $\Omega$
confirms the continuous nature of the WSM-SDW transition. The approximate
Green's function for the lattice model, determined by the stationary
point of $\Omega$, is then used to evaluate the order parameter $\langle M{}_{z}^{(0,0,\pi)}\rangle$
as a function of $U$. Finally, we present the interaction-induced
phase diagram of the WSM model, as a function of the tilt $w_{z}$. 

\begin{figure*}
\begin{centering}
\includegraphics[width=0.8\textwidth]{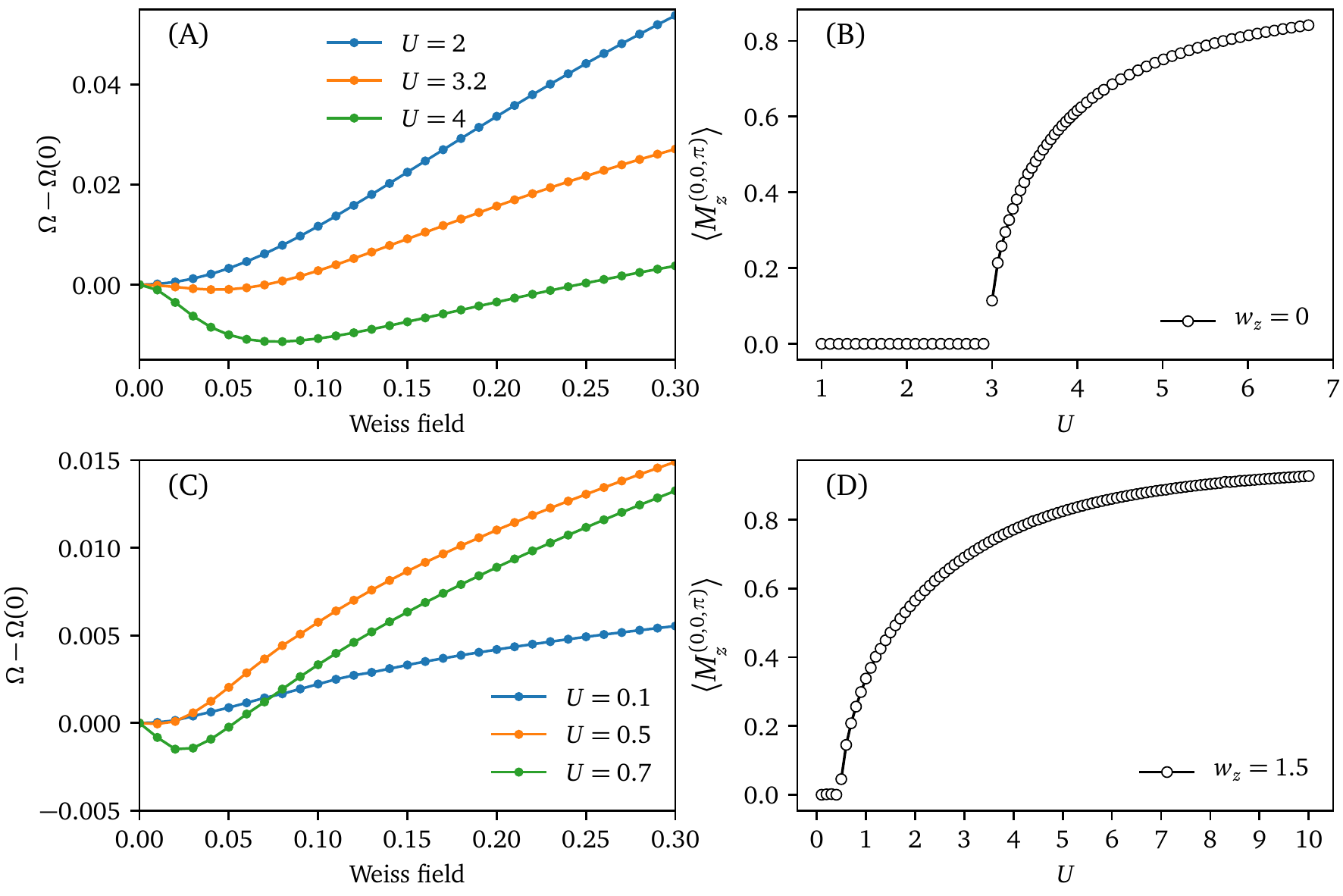}
\par\end{centering}
\caption{\label{fig:sef-vca}Panel (A): Behavior of the Potthoff functional
$\Omega$ as a function of the Weiss field $h$ 
for different values of $U$ (in the vicinity of the critical interaction
$U_{c}$) for the type-I WSM with model parameters $v_{x}=v_{y}=v_{z}=1$
and $w_{z}=0$. Panel (B): Evolution of the SDW order parameter $\langle M{}_{z}^{(0,0,\pi)}\rangle$
as a function of $U$, obtained from our VCA calculations. Panels
(C) and (D): Corresponding results for the type-II WSM model,
with $v_{x}=v_{y}=v_{z}=1$ and $w_{z}=1.5$. We find that for both these models,
the Potthoff functional $\Omega$ develops a minimum at a nonzero
value of the Weiss field $h$ , at a critical
value of the interaction strength, $U=U_{c}$. This is an indication
of a continuous transition to the corresponding ordered state. The
behavior of the order parameter $\langle M{}_{z}^{(0,0,\pi)}\rangle$
as a function of $U$, obtained from the VCA approach, also indicates
a second-order transition at $U=U_{c}$. Note that the value of $U_{c}$
depends sensitively on the tilt parameter $w_{z}$, and is significantly
diminished in the parameter regime where $|w_{z}|>1$.}
\end{figure*}

\subsubsection{Type-I:}

\begin{figure}
\begin{centering}
\includegraphics[width=0.9\columnwidth]{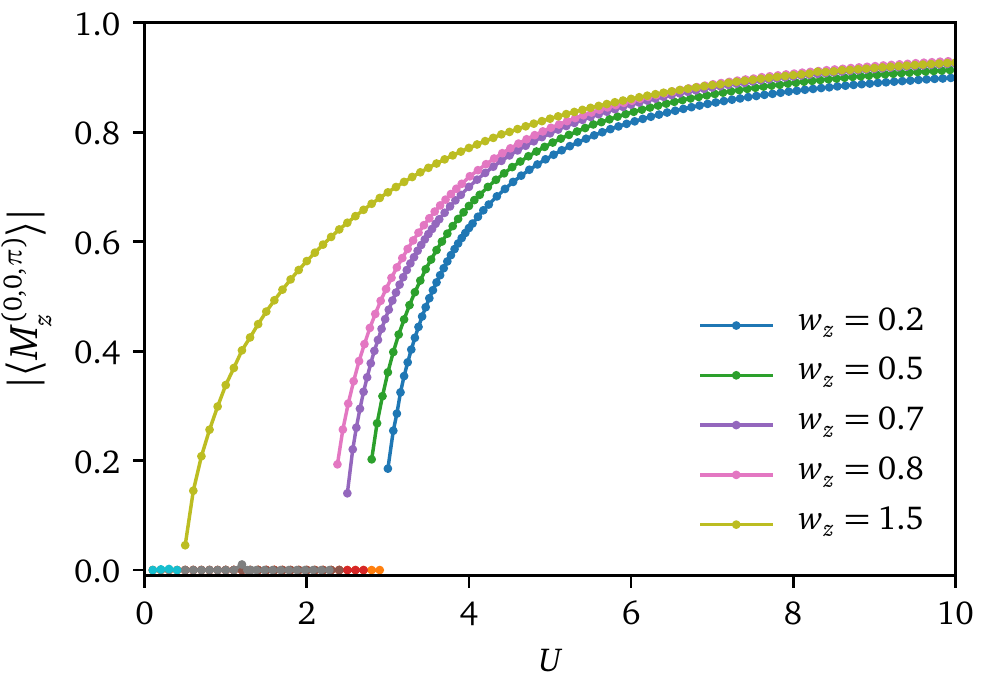}
\par\end{centering}
\caption{\label{fig:diffwz}Evolution of the SDW order parameter $\langle M{}_{z}^{(0,0,\pi)}\rangle$
as a function of the interaction strength $U$, obtained from our
VCA calculations, for different values of the tilt parameter $w_{z}$,
with $v_{x}=v_{y}=v_{z}=1$. Note that as the magnitude of $w_{z}$ increases,
the magnetic transition is found to occur at smaller values of $U=U_{c}$,
which implies that the semimetallic phase is now more susceptible to interactions.
In particular, once in the over-tilted type-II regime, i.e.
$|w_{z}|>|v_{z}|$, we find a sharp decrease in the critical value
of the interaction $U_{c}$, due to the appearance of a finite Fermi
surface with neighboring electron and hole pockets. This change is
also evident from the behavior of the critical interaction $U_{c}$
as a function of $w_{z}$, plotted in Fig.~\ref{fig:phasediag}.}
\end{figure}

\begin{figure}
\begin{centering}
\includegraphics[width=0.8\columnwidth]{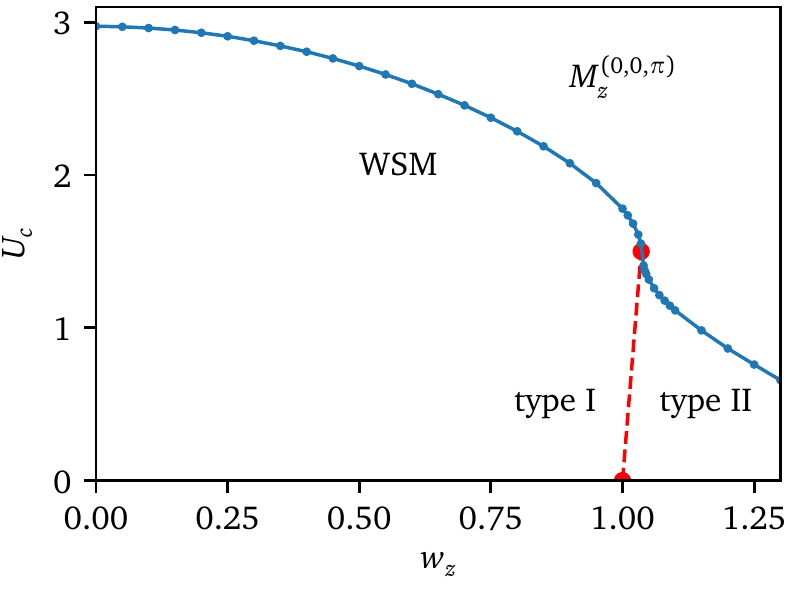}
\par\end{centering}
\caption{\label{fig:phasediag}The interaction-induced phase diagram for the
Weyl semimetal model defined in Eq.~\eqref{eq:1-1}, as a function of
the tilt parameter $w_{z}$ and the corresponding critical interaction
$U_{c}$. The latter determines the phase boundary between the WSM
and $M_{z}^{(0,0,\pi)}$ (SDW) phases for increasing values of $|w_{z}|$.
For increasing values of $w_{z}$, the critical interaction strength
for the transition is found to decrease. We find a singularity in the derivative of $U_{c}$, which
indicates the appearance of a finite Fermi surface in the over-tilted
type-II regime. The dashed red line demarcates the transition between the type-I and type-II regimes. }
\end{figure}

Here, we describe the VCA results for a type-I WSM model with
parameters $v_{x}=v_{y}=v_{z}=1$ and $w_{z}=0$. Fig.~\ref{fig:sef-vca}(A)
shows the behavior of the Potthoff functional $\Omega$ as a function
of the Weiss field $h$  for different values
of $U$. For $U<U_{c}$, $\Omega$ has a single minimum at $h=0$,
corresponding to the gapless semimetallic state. At $U\approx3=U_{c}$,
it develops a new minimum at a finite value of $h$ 
(with a maximum at $h=0$), indicating
a continuous transition to the corresponding magnetically ordered
state. 
We then use the corresponding solution to calculate various physical properties of the model,
and in particular, the order parameter $\langle M{}_{z}^{(0,0,\pi)}\rangle$
as a function of $U$, which becomes nonzero  at the critical value $U=U_{c}$,
as illustrated in Fig.~\ref{fig:sef-vca}(B). The magnitude of $\langle M{}_{z}^{(0,0,\pi)}\rangle$
increases slowly as a function of $U$, and is found to be sensitive
to the band structure parameters.

\subsubsection{Type-II (over-tilted):}

Here, we discuss our VCA results for the WSM model parameters $v_{x}=v_{y}=v_{z}=1$
and $w_{z}=1.5$. We find that the Potthoff functional $\Omega$ once
again develops a minimum at a nonzero value of the Weiss field $h$, with a maximum at $h=0$, 
for a critical interaction strength $U_{c}\sim0.7$ (see Fig.~\ref{fig:sef-vca}(C)),
which is significantly lower than the one obtained for the type-I regime. 
The solution corresponding to the stationary point of $\Omega$
is used to calculate the order parameter $\langle M{}_{z}^{(0,0,\pi)}\rangle$
as a function of $U$ (see Fig.\ref{fig:sef-vca}(D)). Overall, the
behavior of the system in this regime is found to be qualitatively
similar to that of the type-I WSM model, though evidently more
sensitive to interaction effects. This is qualitatively consistent with the CDMFT results
obtained for this model. 

In Fig.~\ref{fig:diffwz}, we use our VCA results for different sets
of band structure parameters to plot the magnitude of the order parameter
$\langle M{}_{z}^{(0,0,\pi)}\rangle$ as a function of $U$, for different
values of $w_{z}$. As the tilt $w_{z}$ is increased, the critical
interaction strength $U_{c}$ decreases. In particular, we find a
significant change in the value of $U_{c}$ when $|w_{z}|>1=|v_{z}|$.
Fig.~\ref{fig:phasediag} shows the interaction-induced phase diagram
for the WSM model defined in Eq.~\eqref{eq:1-1}, as a function of the
parameter $w_{z}$, for $v_{x}=v_{y}=v_{z}=1$. The value of the tilt parameter $w_{z}$, at which the system undergoes a transition from the type-I to the type-II regime (dictated by a singularity in derivative of the critical interaction $U_{c}$) is also found to be slightly renormalized in the presence of interactions, as indicated by the dashed red line in Fig. 8. In general,
for a nontrivial tilt term, the position in the phase diagram where the order appears is highly sensitive to the magnitude of the tilt, and the nature of the SDW order may depend on its direction.

To conclude this part, we find from our VCA analysis that the WSM
model defined by us in Sec.~\ref{sec:model} shows a continuous transition
to an SDW ordered state at a critical value of the interaction strength
$U_{c}$, which is particularly sensitive to the magnitude of the tilt parameter, and this order becomes more robust for increasing values
of $U$. 

\section{\label{sec:conclusions}conclusions}

In summary, we have studied the effect of a local repulsive interaction
$U$ on an inversion-symmetry breaking Weyl semimetal (WSM) model
using the Cluster Dynamical Mean Field Theory (CDMFT) and Variational
Cluster Approximation (VCA) methods. We examine the evolution of the
system as a function of the interaction strength $U$, taking into
account the effect of a nonzero tilt parameter $w_{z}$. We find that
the system undergoes a second-order transition at a critical value
of the interaction $U=U_{c}$, to a spin density wave (SDW) ordered
state, with an ordering wavevector $Q=(0,0,\pi)$ and magnetization
in the $z$-direction. For the untilted dispersion, this is equivalent
to the corresponding state with a wavevector $Q=(\pi,0,0)$ and magnetization
in the $x$-direction. These wavevectors connect Weyl nodes of opposite
chiralities, and such instabilities are favored by the
nesting between these points in the band structure. This result makes sense from a physical point of view, since the Weyl nodes are pinned to high-symmetry points in our model, preventing the movement of nodes of opposite chiralities towards each other. The phase transition
is accompanied by the gradual appearance of a gap in the spectrum.
The Berry flux associated with the Weyl nodes is also found to disappear
in the ordered state due to the backfolding of the Weyl nodes with
opposite chiralities onto each other. 

In the type-II or over-tilted
regime, the corresponding phase transition occurs at a significantly
lower value of $U_{c}$, indicating that the WSM phase is more susceptible
to interactions in this case. The nature of the transition as well
as the magnetic order is confirmed by the results of our VCA calculations.
We then obtain the ground-state phase diagram for the WSM model, as
a function of the tilt parameter $w_{z}$, and find that the critical value of the tilt at which the system undergoes a transition from the type-I to the type-II WSM phase is renormalized in the presence of interactions. Spin-density wave
instabilities have also appeared in previous studies on type-I
and type-II Weyl semimetal models, as well as Dirac semimetal models, using different methods~\citep{PhysRevB.96.165203,PhysRevB.89.235109,PhysRevB.90.035126,PhysRevB.94.125135,PhysRevB.94.241102,PhysRevB.94.075115,PhysRevB.98.165133,PhysRevB.102.125119}, primarily analytical. 

There have been a handful of studies in the literature on interaction
effects in Weyl semimetals which have employed Quantum Cluster Methods~\citep{PhysRevLett.113.136402,PhysRevB.94.241102,Kang2019}, and a prominent example, similar in spirit to our work, is Ref. \onlinecite{PhysRevB.94.241102},
where the VCA approach is used to investigate the effect of both repulsive
and attractive interactions on a time-reversal symmetry breaking type-I
WSM model with tetragonal symmetry, using a slab geometry. Unlike in the model used by us, the specific symmetries  and structure of this model allow the positions of the Weyl nodes to be unrestricted along one of the directions in momentum space. Unlike Ref. \onlinecite{PhysRevB.94.241102}, we also employ the CDMFT approach for our analysis, which is useful for studying a possible Mott insulating phase. As mentioned earlier,
we restrict ourselves to the consideration of repulsive interactions
at half-filling, and more generally, for longer-range or attractive
interactions, one should take into account competing instabilities,
such as charge density wave and superconducting states. 
While the
Quantum Cluster Methods used in this analysis have the advantage of being
nonperturbative and are especially useful in the strongly interacting
limit, they only take into account short-range correlations, and may
therefore overemphasize the order. 

Our treatment may easily be generalized to more complicated Weyl semimetal
models, such as for multi-Weyl systems with quadratic tilt terms, and
such problems will be addressed in future studies. 

\begin{figure*}
\begin{centering}
(a)\includegraphics{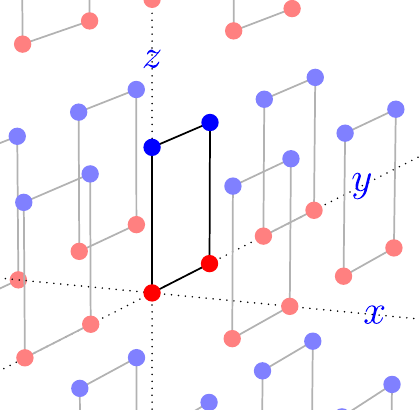}\hspace{1mm}(b)\includegraphics{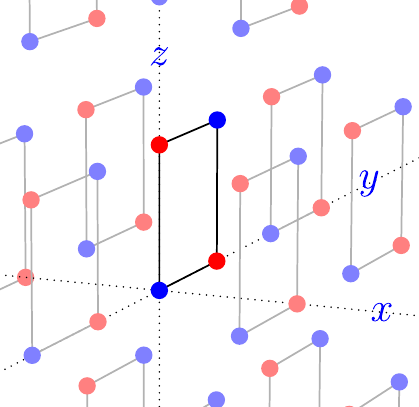}
\par\end{centering}
\caption{\label{fig:qcmclusters}The arrangement of the 4-site clusters employed
for our CDMFT calculations, with superlattice vectors that can accommodate
the SDW orders considered by us, i..e. (a) $M_{z}^{(0,0,\pi)}$ with wave vector
$Q=(0,0,\pi)$ and (b) $M_{z}^{(\pi,\pi,\pi)}$ with wave vector $Q=(\pi,\pi,\pi)$. Here,
we consider a cluster of side $\alpha$, where $\alpha$ is the lattice constant
and the superlattice vectors are $(1,1,0)$, $(0,2,0)$ and $(0,0,2)$,
with distances measured in units of $\alpha$.}
\end{figure*}

\begin{figure*}
\begin{centering}
(a)\includegraphics[scale=0.8]{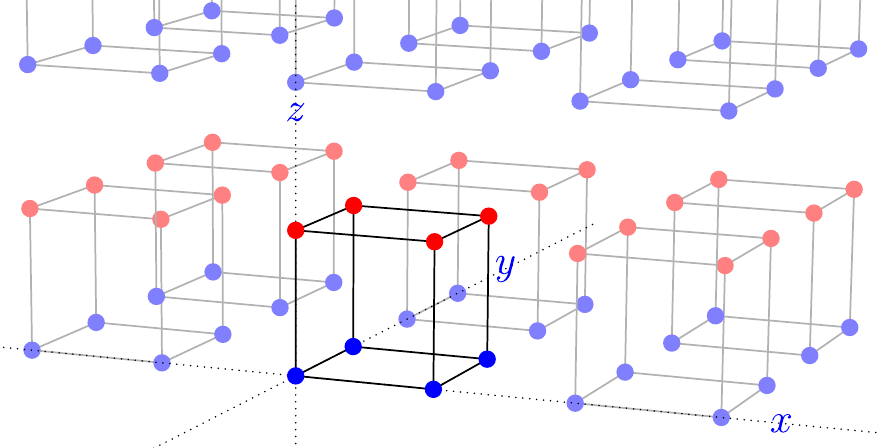}\hspace{1mm}(b)\includegraphics[scale=0.8]{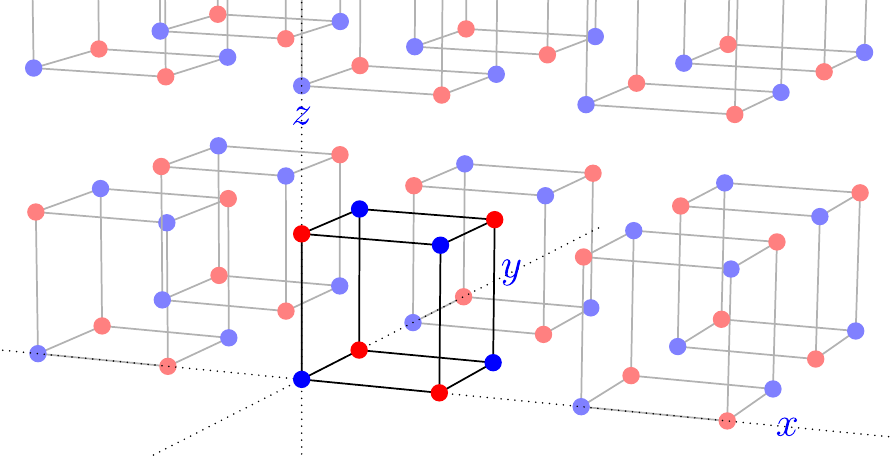}
\par\end{centering}
\caption{\label{fig:qcmclusters1}The 8-site cubic cluster employed by us for our VCA computations with
an illustration of the competing solutions for SDW orders, i.e. (a)
$M_{z}^{(0,0,\pi)}$ with wavevector $Q=(0,0,\pi)$ and (b) $M_{z}^{(\pi,\pi,\pi)}$ with wavevector
$Q=(\pi,\pi,\pi)$. Here, we consider a cluster of side $\alpha$, where
$\alpha$ is the lattice constant, and the superlattice vectors are $(2,0,0)$,
$(0,2,0)$ and $(0,0,2)$, with distances measured in units of $\alpha$. }
\end{figure*}

\begin{acknowledgments}
Computing resources were provided by Compute Canada and Calcul Qu\'ebec.
\end{acknowledgments}


\appendix*
\section{Effective hamiltonian at large $U$}

We use the strong-coupling expansion method to obtain the effective
spin Hamiltonian for our model in the large-$U$ limit. For $w_{z}=0$,
the effective Hamiltonian is given by
\begin{align*}
H_{eff} & =J_{z}\sum_{\langle ij\rangle{}_{z}}(S_{i}^{z}S_{j}^{z}-S_{i}^{x}S_{j}^{x}-S_{i}^{y}S_{j}^{y})\\
 & +J_{x}\sum_{\langle ij\rangle_{x}}(-S_{i}^{z}S_{j}^{z}+S_{i}^{x}S_{j}^{x}-S_{i}^{y}S_{j}^{y})\\
 & +J_{y}\sum_{\langle ij\rangle_{y}}(-S_{i}^{z}S_{j}^{z}-S_{i}^{x}S_{j}^{x}+S_{i}^{y}S_{j}^{y}),
\end{align*}
where $J_{x}=J_{y}=\frac{4v_{x}^{2}}{U}$, $J_{y}=\frac{4v_{y}^{2}}{U}$ and $J_{z}=\frac{4v_{z}^{2}}{U}$
in terms of our WSM model parameters. For $J_{z}\gg J_{x}$, a ferromagnetic
order along $z$ is favored with spins in the $xy$ plane, with an
antiferromagnetic order along $(\pi,0,0)$, as a correction, if $\mathbf{S}$
is along $x$. The same behavior would be observed in the other two
directions if we have $J_{x}\gg J_{z}$, for instance, so the ratios $\frac{J_{z}}{J_{x}}$ or $\frac{J_{z}}{J_{y}}$
do not have a qualitative effect on the behavior of the above Hamiltonian.
This is consistent with the equivalence that we observe, for $w_{z}=0$,
between the SDW order $M_{z}^{(0,0,\pi)}$ (with wavevector $(0,0,\pi)$
and magnetization in the $z$ direction) and $M_{x}^{(\pi,0,0)}$
(with wavevector $(\pi,0,0)$ and magnetization in the $x$-direction).
However, we find no such equivalence to be present in either the noninteracting
or the large-$U$ limit, when $w_{z}\neq0$. 





%



\end{document}